\documentclass[journal=jacsat,layout=twocolum,manuscript=communication]{achemso}

\usepackage{chemformula} % Formula subscripts using \ch{}
\usepackage[T1]{fontenc} % Use modern font encodings

\newcommand\J[2]{\mathbf{J}_{#1}^{\mathrm{#2}}}
\newcommand\X[2]{\mathbf{X}_{#1}^{\mathrm{#2}}}
\newcommand\ONS[2]{\Omega_{#1}^{\mathrm{#2}}}
\newcommand\tn[2]{t_{#1}^{\mathrm{#2}}}
\newcommand\afac[2]{a_{#1}^{\mathrm{#2}}}
\newcommand\Amat[2]{A_{#1}^{\mathrm{#2}}}

\author{Yunqi Shao}%
\affiliation{Department of Chemistry-\AA ngstr\"om Laboratory, Uppsala
  University, Lägerhyddsvägen 1, P. O. Box 538, 75121 Uppsala, Sweden}%

\author{Harish Gudla}%
\affiliation{Department of Chemistry-\AA ngstr\"om Laboratory, Uppsala
  University, Lägerhyddsvägen 1, P. O. Box 538, 75121 Uppsala, Sweden}%

\author{Daniel Brandell}%
\affiliation{Department of Chemistry-\AA ngstr\"om Laboratory, Uppsala
  University, Lägerhyddsvägen 1, P. O. Box 538, 75121 Uppsala, Sweden}

\author{Chao Zhang}%
\email{chao.zhang@kemi.uu.se}%
\affiliation{Department of Chemistry-\AA ngstr\"om Laboratory, Uppsala
  University, Lägerhyddsvägen 1, P. O. Box 538, 75121 Uppsala, Sweden}%

\title[]{Transference number in polymer electrolytes: mind the reference-frame
  gap}


\begin{document}

\begin{tocentry}

  \includegraphics[height=3.5cm]{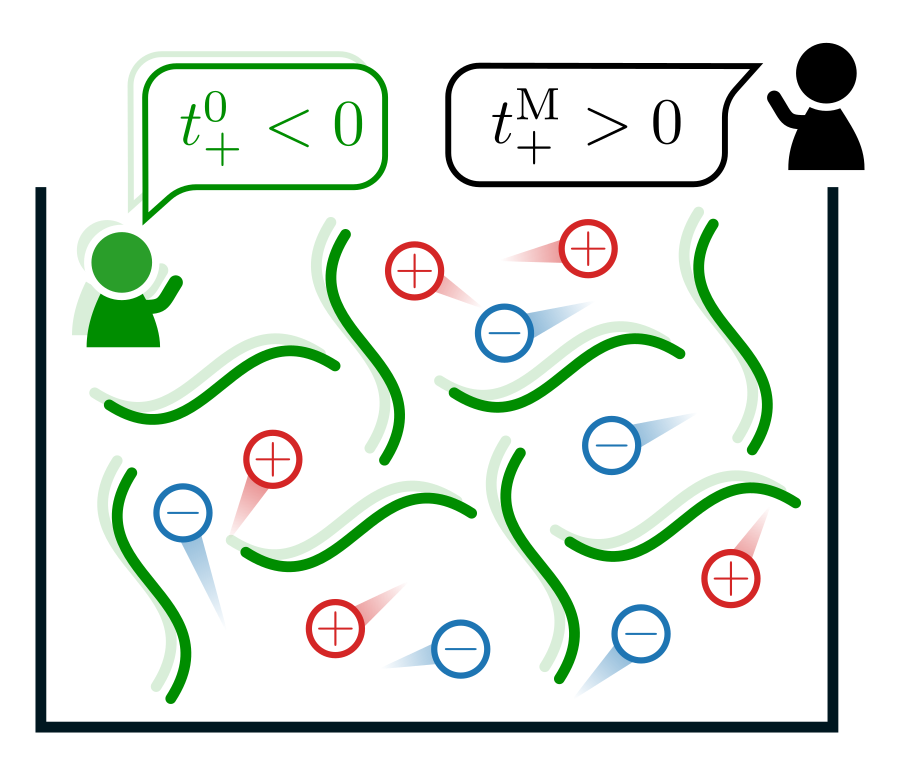}

\end{tocentry}

\begin{abstract}
  The transport coefficients, in particular the transference number, of
  electrolyte solutions are important design parameters for electrochemical
  energy storage devices. Recent observation of negative transference
  numbers in \ch{PEO-LiTFSI} under certain conditions has generated much discussion about its
  molecular origins, by both experimental and theoretical means. However, one overlooked factor in these efforts is the importance of the reference frame (RF). This creates a non-negligible gap when comparing experiment and simulation, because the fluxes in the experimental measurements of transport coefficients and in the linear response theory used in the molecular dynamics simulation are defined in different RFs. In this work, we show that by applying a proper RF transformation, a much improved agreement between experimental and simulation results can be achieved. Moreover, it is revealed that the anion mass and the anion-anion correlation, rather than ion aggregates, play a crucial role for the reported negative transference numbers.
\end{abstract}

% + Introduction
One factor that limits the fast charging and discharging of lithium and lithium-ion
batteries is the build-up of a salt concentration gradient in the cell during operation,
\cite{2018_MindemarkLaceyEtAl,Choo20} since the anion flux due to migration must be
countered by that of diffusion at steady state. It is therefore desirable
for the electrolyte material to carry a greater fraction of cation for
migration, so as to minimize the concentration gradient. This
fraction, known as the cation transference number, is thus of vital importance in the
search for novel electrolyte materials. It is therefore problematic that conventional liquid electrolytes display rather low such numbers, and even more troublesome that they are even lower for solid-state polymer electrolytes based on polyethers.   

While the condition of a uniform concentration when measuring the transference number can be achieved in
typical aqueous electrolytes, its experimental determination in polymer electrolytes is much more
challenging, due to
the continuous growth of the diffusion
layer.\cite{1987_EvansVincentEtAl} At low concentration, the effect of the concentration gradient may be estimated by assuming an ideal solution without ion-ion interactions, as is done in the Bruce-Vincent method
\cite{1987_BruceVincent}. At higher concentrations, its effect on the transference number can be taken into account by the concentrated solution theory developed by Newman, and obtained through a combination of experimental measurements~\cite{2020_NewmanBalsara}.

% - Conversion of transference Number
The cation transference number $\tn{+}{0}$ measured in these experiments is defined typically
in the solvent-fixed reference frame (RF), denoted by the superscript
$\mathrm{0}$ here \cite{1966_Miller}. However, the transference number $\tn{+}{M}$ as computed in molecular dynamics (MD) simulation based on the linear
response theory~\cite{1965_Zwanzig}, is instead related to the velocity correlation functions under the
barycentric RF (denoted by the superscript $\mathrm{M}$). This difference creates a conceptual gap when comparing experiments and simulations, and to interpret results measured in different types of experiments, when seeking the molecular origin behind the observed phenomenon.

To illustrate this point, we here study a typical polymer electrolyte system: \ch{PEO-LiTFSI}. For this, negative $\tn{+}{0}$ has been reported with Newman's approach \cite{2018_VillaluengaPeskoEtAl, 2021_HoffmanShahBalsara}, which has rendered much discussion in the literature \cite{2019_RosenwinkelSchoenhoff, 2018_MolinariMailoaEtAl, 2021_LooFangEtAl}. While the formation of ion aggregates has
often been suggested to cause such negative $\tn{+}{0}$~\cite{2018_MolinariMailoaEtAl}, only marginally negative
values were observed in MD simulations~\cite{2019_FranceLanordGrossman}, even when the correlation due to charged ion clusters was considered explicitly.

To reconcile these observations, we will first investigate how the choice of RF
affect the transference number. In fact, it is possible to
relate $\tn{+}{M}$ to $\tn{+}{0}$  via a simple transformation rule, as shown by Woolf and
Harris\cite{1978_WoolfHarris}:
\begin{equation}
  \label{eq:t_transform}
  \omega_{0}\tn{+}{0} = \tn{+}{M}-\omega_{-}
\end{equation}
where the mass fraction of species $i$ is denoted as $\omega_{i}$. According to
Eq.~\ref{eq:t_transform}, the relation between $\tn{+}{0}$ and $\tn{+}{M}$
depends only on the composition, specifically the mass fractions, of the
electrolyte. 

\begin{figure}
  \centering
   \includegraphics[width=\linewidth]{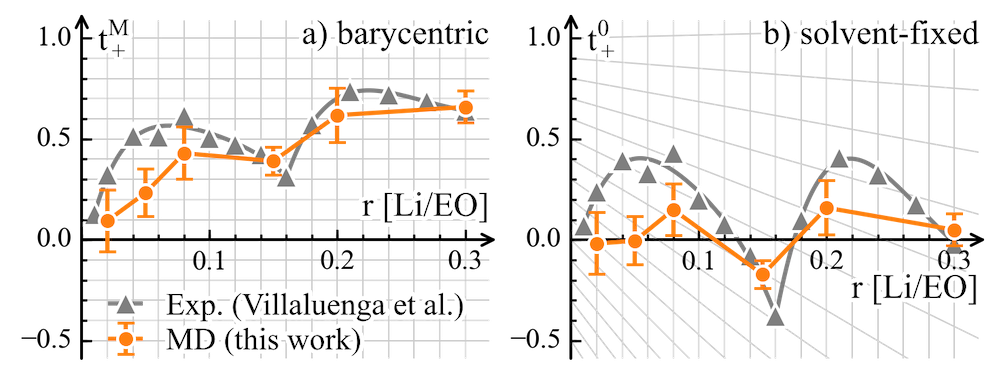}
   \caption{The transference number under a) barycentric RF and b) solvent-fixed
    RF in \ch{PEO-LiTFSI} for different concentrations r [Li/EO] (the ratio of Li:ether oxygen). The conversion rule of $\tn{+}{}$ as determined by
    Eq.~\ref{eq:t_transform} is shown by projecting the grid of a) to b). The
    experimental data and fitting of $\tn{+}{0}$ re reproduced from
    Ref.~\citenum{2018_VillaluengaPeskoEtAl}. The transfer number in MD simulations are
    computed from the corresponding Onsager coefficients using Eq.~\ref{eq:ons2tn}, see the Supporting
    Information for simulation details.}
  \label{fig:t_transform}
\end{figure}
% - Exp vs. MD, negative transference number
While the two
transference numbers are equivalent at the limit of infinite dilution ($\omega_{0} \to 1$), they become distinctly different at higher concentration. As shown
in Fig.~\ref{fig:t_transform}, at the concentration where negative $\tn{+}{0}$
is observed, $\tn{+}{M}$ is still positive. Moreover, $\tn{+}{}$ generally  shifts downward in the solvent-fixed RF as the concentration increases, as seen in Fig.~\ref{fig:t_transform}. This trend
can be expected, since at the other limit ($\omega_{0} \to 0$), $\tn{+}{M}$ must converge
to the $\omega_{-}$ in order to satisfy Eq.~\ref{eq:t_transform}. This suggests that $\tn{+}{0}$ will become increasingly sensitive at higher concentration since its value will be determined by the motion of a small fraction of solvent molecules.

% + Onsager coefficients - Necessity for interpretation
The distinction between  $\tn{+}{M}$ and  $\tn{+}{0}$ may already explain why negative transference number is seldom
observed in MD simulations, where the barycentric RF is the default setting. But more
importantly, the strong dependency of $\tn{+}{}$ on the RF suggests that the intuitive explanation of the observed negative $\tn{+}{0}$ being due to the population of ion aggregates is not necessarily the case. Instead, as pointed out in
recent studies\cite{VargasBarbosa:2020eg, zhang2020, 2021_PfeiferAckermannEtAl, 2021_FongSelfEtAl,gudla21}, the explicit
consideration of ion-ion correlations is essential to understand ion transport
in polymer electrolytes.

% - Setup, computational details
In the following, we will show how the ion-ion correlations contribute to the negative transference number in light of the RF. In the Onsager phenomenological equations~\cite{onsager45}, the flux $\J{i}{S}$ of species under
a reference frame $\mathrm{S}$ can be considered as the linear response of the
external driving forces $\X{j}{}$ acting on any species $j$:

\begin{equation}
  \label{eq:onsager}
  \J{i}{S} = \sum_{j} \ONS{ij}{S} \X{j}{}
\end{equation}

where $\ONS{ij}{S}$ are the Onsager coefficients. For the index $j$, here we denote the
solvent as $0$, the cation as $+$, and the anion as $-$. In addition, the fluxes satisfy the following RF condition   $\sum_{i}\afac{i}{S}\J{i}{S} =0 $, where
$\afac{i}{S}$ are the proper weighing factors, i.e. $\afac{i}{M}=M_{i}$ for the
barycentric RF and $\afac{i}{0}=\delta_{i0}$ for the solvent-fixed RF.  Then, a unique set of the Onsager coefficients can be determined by applying the Onsager
reciprocal relation: $\ONS{ij}{S}=\ONS{ji}{S}$, and the RF
constraint $\sum_{i}a_{i}^{S}\ONS{ij}{S}=0\ \forall j$.

Knowing these Onsager coefficients, one can express the transport properties of interest here, i.e. the transference number and the ionic
conductivity, as:

\begin{align}
  \tn{i}{S} &= \frac{\sum_{j} q_{i}q_{j}\ONS{ij}{S}}{\sum_{j,k} q_{j}q_{k}\ONS{jk}{S}} \label{eq:ons2tn}\\
  \sigma &= \sum_{i,j} q_{i}q_{j}N_{\mathrm{A}}^{2}\ONS{ij}{S}
  \label{eq:conduc}
\end{align}

% - Non-uniqueness
where $q_{i}$ is the formal charge of $i$ and $N_{\mathrm{A}}$ is the Avogadro
constant. It is worth noting that unlike the transference number, the ionic conductivity is RF-independent because of the charge neutrality condition.  

% - Transformation rules
\begin{figure}
  \centering
  \includegraphics[width=\linewidth]{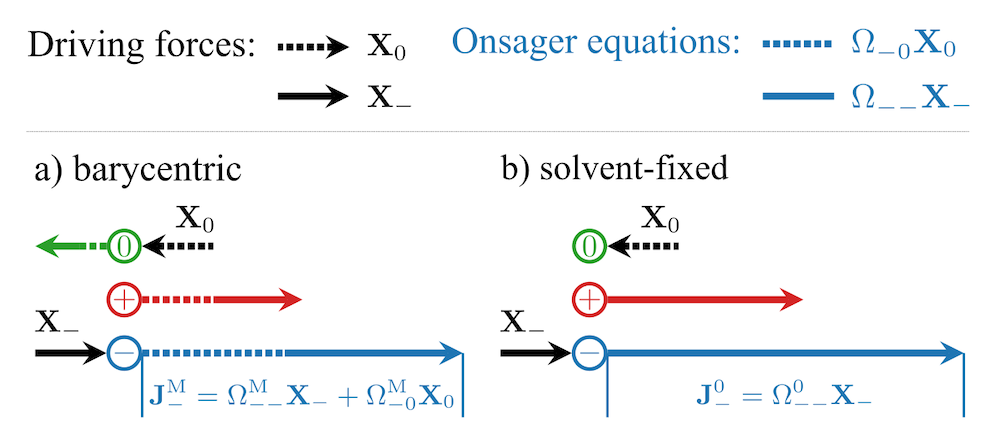}
  \caption{An illustration of the transformation procedure when converting
    $\ONS{ij}{M}$ to $\ONS{ij}{0}$ for the case where the driving force acting
    on cation is zero. The dashed lines indicate relevant parts related to the solvent.}
  \label{fig:ill_transform}
\end{figure}

While the transformation of $\tn{+}{}$ from the solvent-fixed RF to the barycentric RF can follow the straightforward rule of Eq.~\ref{eq:t_transform}, the corresponding RF transformation of $\ONS{ij}{}$ is not trivial. This is illustrated by a simplified example shown in Fig~\ref{fig:ill_transform}, where driving force acting
on the cation is assumed to be zero. In the barycentric RF both driving forces $\X{0}{}$ acting on solvent and $\X{-}{}$ acting on anion will contribute to the anion flux $\J{-}{M}$.  When transforming the Onsager coefficients to the solvent-fixed RF, only the driving force $\X{-}{}$ contributes to the anion flux $\J{-}{0}$, as $\ONS{-0}{0}=0$ by construction.

Nevertheless, the general transformation rule can be derived using the independent fluxes and driving forces~\cite{1960_KirkwoodBaldwinEtAl}, which is consistent with the above
constructions. Following the notation of Miller \cite{1986_Miller}, one can consider
only the $n-1$ independent fluxes and driving forces in a $n$ component system,  where the flux of the solvent $\J{0}{}$ is treated as a redundant variable. This leads to the following set of rules for the RF transformation:

\begin{align}
  \Amat{ij}{RS} &=\delta_{ij}+\frac{c_{i}}{\sum_{k}\afac{k}{R}c_{k}}
                  \left(\frac{\afac{0}{R}\afac{j}{S}}{\afac{0}{S}}-\afac{j}{R}\right)\\
  \J{i}{R} &= \sum_{j\neq 0}\Amat{ij}{RS} \J{j}{S}\\
  \ONS{ij}{R} &= \sum_{k,l \neq 0}\Amat{ik}{RS} \ONS{kl}{S} \Amat{jl}{RS}
                \label{eq:ons_transform}
\end{align}

where $\Amat{ij}{RS}$ is the matrix that converts the independent fluxes from
the reference frame S to R and $c_i$ is the molar concentration of species $i$. The coefficients $\ONS{i0}{R}$ may then be fixed
according to the RF constraint. The specific transformation equations for the
barycentric and solvent-fixed RFs are provided in the Supporting Information.

% - Comparison of the conversion, exp vs. md
This transformation provides the connection between $\ONS{ij}{0}$ measured
experimentally and $\ONS{ij}{M}$ derived from MD simulations. Thus, one can compare Onsager coefficients under a common RF to see whether the simulation describes the same transport mechanism as in experiment or not. Here we computed Onsager coefficients following Miller's derivation
\cite{1966_Miller}, with experimental measurements by Villaluenga et al.
\cite{2018_VillaluengaPeskoEtAl} MD simulations were performed using GROMACS~\cite{2015_AbrahamMurtolaEtAl} and the General AMBER Force
Field\cite{2004_WangWolfEtAl}, from which Onsager coefficients were derived with an
in-house analysis software. Details of the conversion and simulation procedure
can be found in the Supporting Information. In addition, we shall note here that an
alternative set of transport coefficients, i.e. the Maxwell-Stefan diffusion
coefficients, were orginally reported from experiment
\cite{2018_VillaluengaPeskoEtAl}, and they are consistent with the present framework (see the Supporting Information for the inter-conversion). However, the Onsager coefficients are favoured here
because they are well-behaved at any given concentration and therefore helpful to understand the
RF dependency of the ion-ion correlations.

\begin{figure} [ht]
  \centering
  \includegraphics[width=\linewidth]{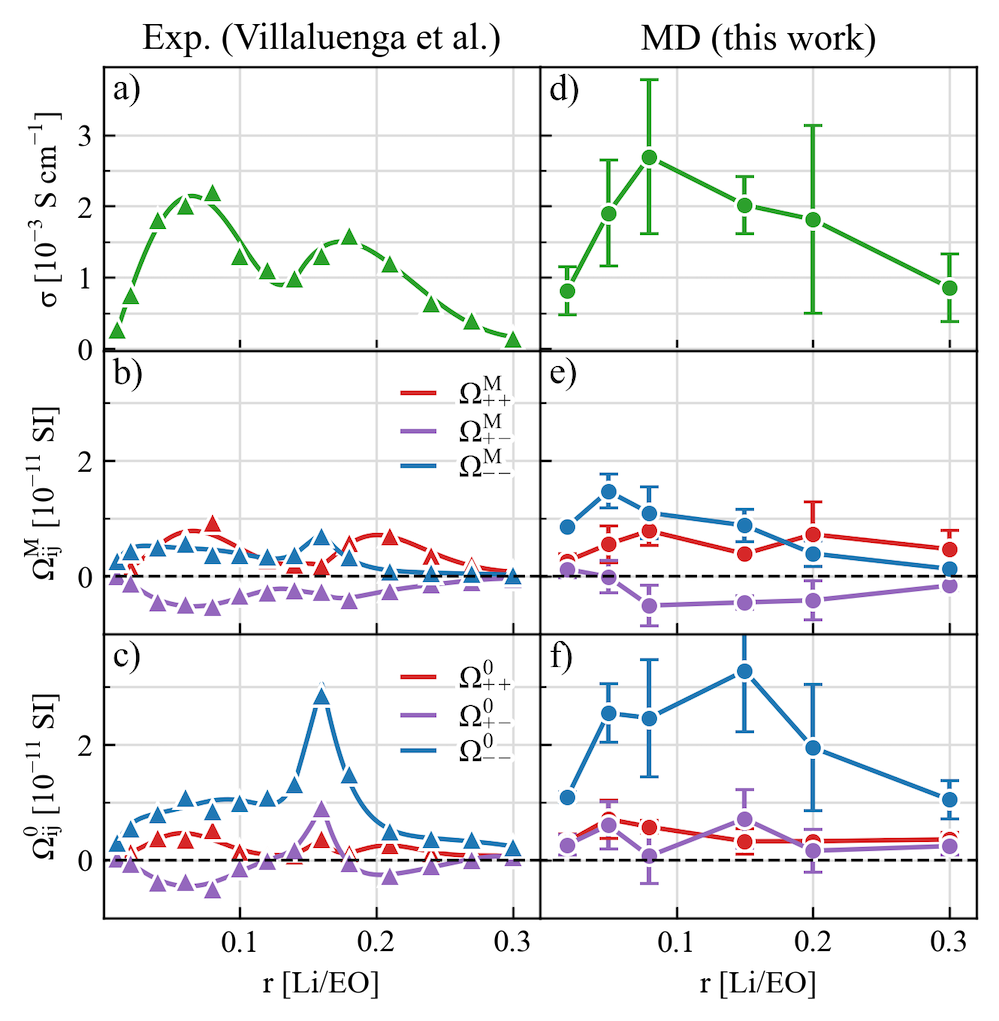}
   \caption{Ionic conductivity and Onsager coefficients under the barycentric and
    solvent-fixed derived from a-c) experimental measurements and d-f) MD
    simulations. The experimental measurements (triangle dots) and fittings
    (curved lines) are taken from Ref.~\citenum{2018_VillaluengaPeskoEtAl}. The
    MD simulation results are computed by fitting the mean displacement
    correlations functions, as detailed in the Supporting Information. }
  \label{fig:ons_compare}
\end{figure}

As shown in Fig.~\ref{fig:ons_compare}, the conductivity and Onsager
coefficients obtained from MD simulations generally matches the experimental
values. In particular, $\ONS{+-}{M}$ is negative in the entire concentration range and this indicates an anti-correlation between cations and anions.
Furthermore, we see that the experimentally observed negative transference number at $r=0.15$
is reproduced in the MD simulation, with consistent features of $\ONS{ij}{}$,
namely, $\ONS{--}{0}>\ONS{+-}{0}>\ONS{++}{0}>0$ and
$\ONS{--}{M}>\ONS{++}{M}>0>\ONS{+-}{M}$. These results demonstrate that the experimentally observed negative transference number in PEO-LiTFSI systems is captured with the present force field parameterization used in the MD simulations.

If looking at the effects of RF, we see that $\ONS{--}{}$ and
$\ONS{+-}{}$ changes more significantly upon RF transformation, as compared to
$\ONS{++}{}$. Especially, at $r=0.15$, $\ONS{+-}{M}$ is negative while
$\ONS{+-}{0}$ is positive. This means that the driving force applied to the cations
correlates to a co-directional anion flux in the solvent-fixed RF, but that an opposite
anion flux is found in the barycentric RF. This, together with the observations made above, cannot be
explained by any distribution of ideal charge carrying clusters.

% + Interpretation - Transform rules
To better understand  the underlying physical account, we can look into the Onsager coefficients from a microscopic point of view, as they are related to the
correlations functions of the fluxes. From the equations shown below, it is clear that the RF transformation is equivalent to
transforming either the current-correlation function shown in Eq.~\ref{eq:ons_jcf} or, equivalently,
the displacements of ions shown in Eq.~\ref{eq:ons_mdc}. Thus, this result (Eq.~\ref{eq:ons_mdc_transform}) is
consistent with Eq.~\ref{eq:ons_transform}, and Wheeler and Newman's expression
for $\ONS{ij}{0}$ \cite{2004_WheelerNewman}.

\begin{align}
  \ONS{ij}{0}  = &\frac{\beta}{3}\int d\mathbf{r}\int_{0}^{\infty}
                 dt\left\langle\J{i}{0}(0,0)\cdot\J{j}{0}(\mathbf{r},t))\right\rangle \label{eq:ons_jcf}\\
 = & \lim_{t\to\infty}\frac{\beta}{6VN_\mathrm{A}^2t}
                 \left\langle \Delta\mathbf{r}_{i}^\mathrm{0}(t)\cdot
                 \Delta\mathbf{r}_{j}^\mathrm{0}(t)\right\rangle \label{eq:ons_mdc}\\
            = & \lim_{t\to\infty}\frac{\beta}{6VN_\mathrm{A}^2t}
                 \left\langle \left(\sum_k \Amat{ik}{0M} \Delta\mathbf{r}_{k}^\mathrm{M}(t)\right)
                 \cdot
                 \left(\sum_{l} \Amat{jl}{0M} \Delta\mathbf{r}_{l}^\mathrm{M}(t) \right)
                 \right\rangle  \label{eq:ons_mdc_transform}
\end{align}

where $\beta=1/(k_{\mathrm{B}}T)$ is the inverse temperature, $N_\mathrm{A}$ is the Avogadro constant and $\Delta\mathbf{r}_{i}^\mathrm{R}(t)$ is the total displacement of species $i$ over 
a time interval $t$.

\begin{figure}
  \centering
  \includegraphics[width=\linewidth]{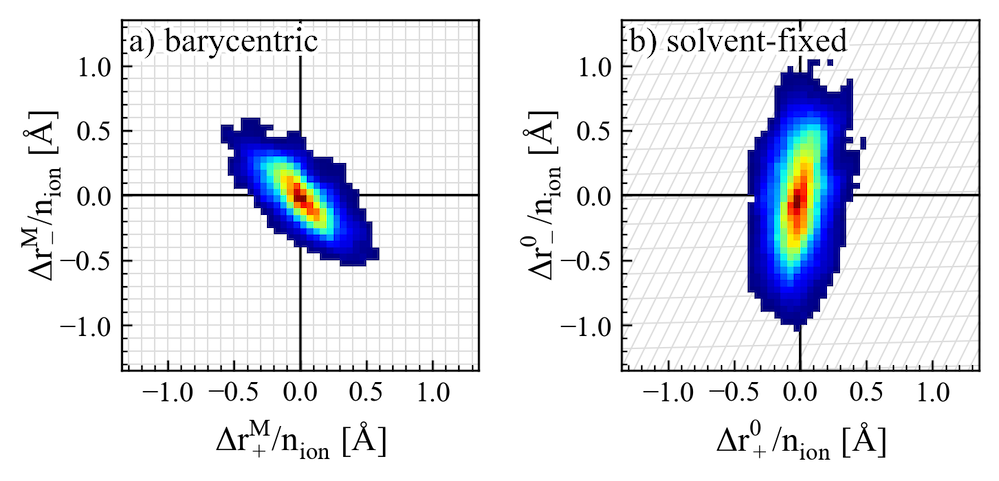}
  \caption{Transformation of the normalized displacement correlations upon a change of
    reference frame. $\Delta r^\mathrm{M}/n_\mathrm{ion}$ is the total displacement $\Delta r^\mathrm{M}$ (of cations ``+" or anions ``-'') normalized by the number of ions $n_\mathrm{ion}$. The correlation is obtained from a 400 ns MD trajectory,
    where the correlation between mean displacements of cations and anions over
    $\Delta t$=10ns is plotted in a) the barycentric RF and b) the solvent-fixed
    RF. The RF transformation according to Eq.~\ref{eq:ons_transform} is
    visualized as the projection of grid lines from a) to b).}
  \label{fig:ons_transform}
\end{figure}

Based on this result, the conversion of Onsager coefficients upon an RF transformation can
be visualized as an affine transformation of ion displacement, as shown in
Fig.~\ref{fig:ons_transform}. At $r=0.15$, the displacement of cations and anions
are apparently anti-correlated in the barycentric RF, while the correlation
becomes positive in the solvent-fixed RF. This can be rationalized, since the
motion of anions in the barycentric RF entails the motion of solvent in the
opposite direction, giving rise to the enhanced anion motion and the positive cation-anion correlation in the
solvent-fixed RF. On the other hand, the motion of
cations induces a much less significant effect, as signified by the small
distortion along the x-axis. This points in the direction that anions play a significant role for the transference number of Li$^+$, not only by its relative motion to the cation. 

\begin{figure}
  \centering
  \includegraphics[width=\linewidth]{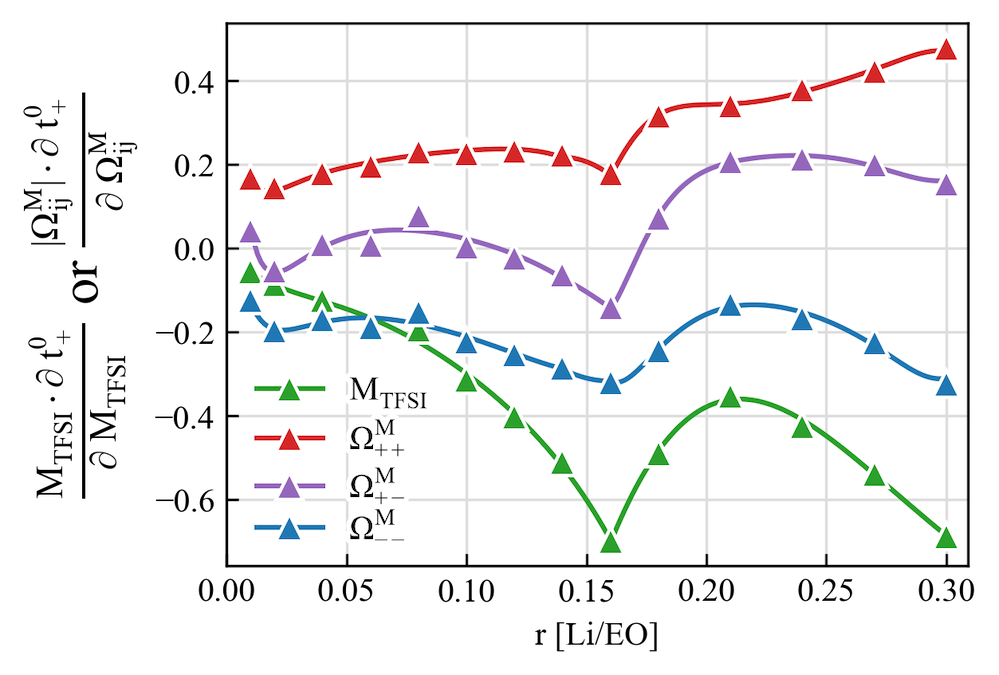}
  \caption{The sensitivity of transference number $\tn{+}{0}$  in solvent-fixed RF
    to the variations in the anion molecular weight $\mathrm{M}_{\ch{TFSI}}$ and
    different Onsager coefficients $\ONS{ij}{M}$ in the barycentric RF. The
    analysis is done by evaluating the partial derivative of $\tn{+}{0}$ to the
    logarithm of $\mathrm{M}_{\ch{TFSI}}$ or $|\ONS{ij}{M}|$, with data derived
    from experimental measurements in Ref.~\citenum{2018_VillaluengaPeskoEtAl}.
    Note that $\ONS{+-}{M}$ is mostly negative as shown in
    Fig.~\ref{fig:ons_compare}, while the other variables are positive.}
  \label{fig:t_sensitivity}
\end{figure}

Indeed, the sign of the experimentally measured $\tn{+}{0}$ depends not only on
$\ONS{++}{M}-\ONS{+-}{M}$, as is evident in Eq.~\ref{eq:ons2tn}, but also on the
$\ONS{--}{M}$ and the anion mass fraction. The importance of the anion-anion
correlation and the anion mass is demonstrated in
Fig.~\ref{fig:t_sensitivity}, where the partial derivative of $\tn{0}{+}$ shows
its strong dependency on the anion mass and Onsager coefficients. An increase of the
anion mass introduces an even stronger reduction of the transference number $\tn{0}{+}$, and therefore $\tn{0}{+}$ is more likely to be negative. The same effect occurs when the anion-anion correlation becomes stronger and $\ONS{--}{M}$ becomes larger.  This suggests a direct connection between the observed negative $\tn{+}{0}$ and a strong anion-anion correlation found at higher concentrations. The latter effect was also indicated in a recent X-ray scattering study of PEO-LiTFSI systems \cite{2021_LooFangEtAl}.

In summary, our present analysis reveals a strong RF dependency of the
transference number and the Onsager coefficients in the \ch{PEO-LiTFSI} system.
With a proper transformation, the Onsager coefficients can be used as a
rigorous test to compare the transport properties from experimental measurements
and MD simulations, as shown here. This will provide a new ground to refine force field parameterization, for example, by including the subtle effects of electronic polarization~\cite{Borodin:2009jb}, although we found that the standard force field already captures the main features observed in experiments.

Not only do our results demonstrate that the experimentally observed negative
$\tn{+}{0}$ can be reproduced with MD simulations, but they also show that cations and anions are anti-correlated in the barycentric RF ($\ONS{+-}{M}$<0) throughout the entire concentration range in both experiment and simulation. While this does not rule out the possibility of short-lived ion aggregates, neither does it support a transport mechanism based on negatively-charged ion clusters. Instead, we show that a large anion mass
and strong anion-anion correlations can be responsible for a negative transference number of  $\tn{+}{0}$. 

Furthermore, the RF-dependence of ion-ion correlations suggests that any discussions about ion-ion correlations need to be done within the same RF. This may shed light on why a different observation was made regarding the sign of $\tn{+}{}$ with alternative experimental approaches such as electrophoretic NMR (eNMR)~\cite{2019_RosenwinkelSchoenhoff}. 

Although we do not expect that all discrepancies in transport properties between different experimental approaches and between experiment and simulation can be resolved by the present analysis, insights regarding the RF dependency of ion-ion correlations, and
direct comparison of the complete set of Onsager coefficients between experiment
and simulation as demonstrated in this work would be essential to elucidate the
ion transport mechanism in polymer electrolytes and alike concentrated
electrolyte systems.

\begin{acknowledgement}
This work has been supported by the European Research Council (ERC), grant no. 771777 “FUN POLYSTORE” and the Swedish Research Council (VR), grant no.  2019-05012. The authors thanks funding from the Swedish National Strategic e-Science program eSSENCE, STandUP for Energy and BASE (Batteries Sweden). The simulations were performed on the resources provided by the Swedish National Infrastructure for Computing (SNIC) at PDC.
\end{acknowledgement}

\begin{suppinfo}
  Details of MD simulations and force field parameters; Computation and conversion of Onsager coefficients in different RFs; Conversion between different sets of transport equations; List of symbols.
\end{suppinfo}

\providecommand{\latin}[1]{#1}
\makeatletter
\providecommand{\doi}
  {\begingroup\let\do\@makeother\dospecials
  \catcode`\{=1 \catcode`\}=2 \doi@aux}
\providecommand{\doi@aux}[1]{\endgroup\texttt{#1}}
\makeatother
\providecommand*\mcitethebibliography{\thebibliography}
\csname @ifundefined\endcsname{endmcitethebibliography}
  {\let\endmcitethebibliography\endthebibliography}{}

%\bibliography{./references.bib}

\end{document}

\makeatletter 
\renewcommand{\theequation}{S\@arabic\c@equation} 

\clearpage

\section{MD simulations and force field parameters}
The General AMBER force field (GAFF)\cite{2004_WangWolfEtAl} parameters were used for describing bonding and non-bonding interactions in PEO and LiTFSI. The force field parameters along with atomic partial charges were obtained and assigned using ACPYPE\cite{SousaDaSilva2012} and ANTECHAMBER\cite{Wang2006} tools. These partial charges on the salt were scaled by a factor of 0.75 to effectively introduce electronic polarizations and also previously shown to better reproduce experimental diffusivities. 

\begin{table}[h!]
\centering
 \begin{tabular}{||c c c c||} 
 \hline
 \multicolumn{2}{||c}{PEO} & \multicolumn{2}{c||}{LiTFSI}\\
 \hline
 Atom type & Charge (e) & Atom type & Charge (e) \\ [0.5ex] 
 \hline\hline
 C & 0.130 & C & 0.417 \\ 
 O$_\textrm{chain}$ & -0.429 & S & 1.113, 1.261 \\
 H$_\textrm{chain}$ & 0.043 & F & -0.204 \\
 O$_\textrm{end}$ & -0.612 & N & -0.794 \\
 H$_\textrm{end}$ & 0.405 & O & -0.485 \\
  &  & Li & 0.750 \\ [1ex] 
 \hline
 \end{tabular}
 \caption{The atomic partial charges of PEO and LiTFSI from GAFF.}
\label{tab:gaff_charges}
\end{table}

The initial polymer MD simulation boxes comprising 200 hydroxyl-terminated poly(ethylene oxide) (PEO) chains, each with 25 monomer units (1.11 kg/mol) were constructed using PACKMOL package\cite{Martinez2009}. Six different salt concentrations were obtained by adding 100, 250, 400, 750, 1000 and 1500 lithium bis(trifluoromethane)sulfonimide (LiTFSI) molecules, corresponding to a [Li/EO] concentration ratio of 0.02, 0.05, 0.08, 0.15, 0.20, and 0.3, respectively. After the energy minimization step, all systems were equilibrated first with NVT (constant number of particles, volume, and temperature) and then NPT (constant number of particles, pressure, and temperature) runs using Bussi-Donadio-Parrinello\cite{Bussi2007} thermostat and a Parrinello-Rahman barostat\cite{Parrinello1981} at 400 K and 1 bar with a time step of 1 fs. The thermostat and barostat coupling constants were set to 0.1 and 2.0 ps, respectively. Then, NPT production runs were carried out for additional 400 ns at 430K to ensure that the Li-ion dynamics has reached diffusive regime and trajectories were saved every 5 ps.

\newpage
\section{Computation and conversion of Onsager coefficients in different RFs}

The Onsager coefficients from MD simulations are computed with the mean
displacement correlations (in the barycentric RF):
\begin{align}
  \ONS{ij}{M}  = \lim_{t\to\infty}\frac{\beta}{6VN_\mathrm{A}^2t}
                 \left\langle \Delta\mathbf{r}_{i}^\mathrm{M}(t)\cdot
                 \Delta\mathbf{r}_{j}^\mathrm{M}(t)\right\rangle
\end{align}

where $\beta$ is the inverse temperature, $N_\mathrm{A}$ is the Avogadro
constant and $\Delta\mathbf{r}_{i}^\mathrm{M}(t)$ is the total displacement of
species $i$ over a time interval $t$. The limit is estimated by fitting the
correlation
$\left\langle \Delta\mathbf{r}_{i}^\mathrm{M}(t)\cdot\Delta\mathbf{r}_{j}^\mathrm{M}(t)\right\rangle$
as a linear function of $t$ over the interval 10-20ns. The MD trajectories were
split into segments of 100ns, and the limit was estimated separately for each
segment, where the standard deviation across segments is taken as the error
estimation in the main text.

The experimental values of Onsager coefficients for an electrolyte solution of the molar concentration $c$ are computed from the measured
conductivity $\sigma$, solvent-fixed cation transference number $\tn{+}{0}$, salt
diffusion coefficients $D_{\mathrm{salt}}$,  and the so-called thermodynamic factor $\tfac$:
\begin{equation}
  \label{eq:exp2ONS0}
  \ONS{ij}{0} =\frac{\tn{i}{0}\tn{j}{0}\sigma}{z_{i}z_{j}F^{2}} + \frac{cD_{\mathrm{salt}}}{2RT \left(\tfac\right)}
\end{equation}

which is equivalent to that derived by Miller\cite{1966_Miller}. A special form of Eq.~7 in the main text is used to convert the Onsager
coefficients for 1:1 electrolytes to solve $\ONS{ij}{M}$ from $\ONS{ij}{0}$ or
vice versa. Written as the linear relation between independent
sets of $\ONS{ij}{}$, one have:
\begin{equation}
\begin{bmatrix}
\ONS{++}{0} \\
\ONS{+-}{0} \\
\ONS{--}{0}
\end{bmatrix} = \frac{1}{\rho_0^2}
\begin{bmatrix}
(\rho_0+\rho_+)^2 & 2\rho_-(\rho_0+\rho_+) & \rho_-^2 \\
\rho_+(\rho_0+\rho_+) & \rho_0^2+\rho_0\rho_++\rho_0\rho_-+2\rho_+\rho_- & \rho_-(\rho_0+\rho_-) \\
\rho_+^2 & 2\rho_+(\rho_0+\rho_-) & (\rho_0+\rho_-)^2
\end{bmatrix}
\begin{bmatrix}
\ONS{++}{M} \\
\ONS{+-}{M} \\
\ONS{--}{M}
\end{bmatrix}
\end{equation}

where $\rho_{0}$, $\rho_{+}$, $\rho_{-}$ are the mass concentration of the solvent,
the cation, and the anion.

\newpage
\section{Conversion between different sets of transport equations}

In the Maxwell-Stefan equations\cite{1997_KrishnaWesselingh}, the driving force is related to relative
velocities between different species through the Maxwell-Stefan diffusion
coefficients $\mathfrak{D}_{ij}$ or the friction coefficients $K_{ij}$:
\begin{align}
  \label{eq:MS_D}
  c_i\nabla\mu_i &= RT\sum_{j\neq i} \frac{c_ic_j}{c_T\mathfrak{D}_{ij}}(\mathbf{v}_j - \mathbf{v}_i) \\
  \label{eq:MS_K}
  c_i\nabla\mu_i &= \sum_{j\neq i} K_{ij}(\mathbf{v}_j - \mathbf{v}_i)
\end{align}

where $c_T=\sum_{i} c_{i}$. It follows that $K_{ij}$ and $\mathfrak{D}_{ij}$ are related as:
\begin{equation}
  \frac{RT}{\mathfrak{D}_{ij}} = \frac{c_{T}}{c_{i}c_{j}}K_{ij}
\end{equation}

The relation between the Maxwell-Stefan coefficients with the Onsager
coefficients can be shown through the modified form of Maxwell-Stefan equations introduced
by Newman \cite{2004_WheelerNewman}, where the relative velocities are replaced
with the velocities in the solvent-fixed RF:
\begin{equation}
    \label{eq:MS_Newman}
    c_i\nabla\mu_i = \sum_{j\ne 0} M_{ij}(\mathbf{v}_j - \mathbf{v}_0)
\end{equation}

$M_{ij}$ are related to $K_{ij}$ as:
\begin{equation}
\label{eq:M_to_K}
\begin{bmatrix}
K_{+0} \\
K_{+-} \\
K_{-0}
\end{bmatrix} =
\begin{bmatrix}
-1 & -1 & 0 \\
0 & 1 & 0 \\
0 & -1 & -1
\end{bmatrix}
\begin{bmatrix}
M_{++} \\
M_{+-} \\
M_{--}
\end{bmatrix}
\end{equation}

Inspecting Eq.~\ref{eq:MS_Newman} and the Onsager equations in the solvent-fixed
RF (note that $\ONS{i0}{0}=0$):
\begin{align}
    \J{i}{0} = c_i(\mathbf{v}_j - \mathbf{v}_0) = -\sum_{j\ne 0} \ONS{ij}{0} \nabla \mu_{j}
\end{align}

It is clear that the $M_{ij}$ and $\ONS{ij}{0}$ are related with the matrix
inversion of transport coefficients, for 1:1 electrolyte solution with concentration $c$:
\begin{equation}
  \begin{bmatrix}M_{++} & M_{+-} \\  M_{+-} & M_{--}\end{bmatrix} = -c^{2}
  \begin{bmatrix}\ONS{++}{0} & \ONS{+-}{0} \\  \ONS{+-}{0} & \ONS{--}{0}\end{bmatrix}^{-1}
\end{equation}

or:
\begin{equation}
\label{eq:Omega0_to_M}
\begin{bmatrix}
M_{++} \\
M_{+-} \\
M_{--}
\end{bmatrix} = \frac{c^{2}}{\detO}
\begin{bmatrix}
-\ONS{--}{0} \\
 \ONS{+-}{0} \\
-\ONS{++}{0}
\end{bmatrix}
\end{equation}

with:
\begin{align}
  \detO &= \detOex \\
        &= \frac{c\sigma D_{\mathrm{salt}}}{2RTF^{2}\left(\tfac\right)}
\end{align}

The consistency between the different sets of transport coefficients may be shown by
representing $\mathfrak{D}_{ij}$ in terms of $\ONS{ij}{0}$:
\begin{align}
  \mathfrak{D}_{+0} &= \frac{c_{0}RT}{c_{T}c}\cdot\frac{\detO}{\ONS{--}{0}-\ONS{+-}{0}}
  \label{eq:ONS02DMS1}\\
  \mathfrak{D}_{+-} &= \frac{RT}{c_{T}}\cdot\frac{\detO}{\ONS{+-}{0}}
  \label{eq:ONS02DMS2}\\
  \mathfrak{D}_{-0} &= \frac{c_{0}RT}{c_{T}c}\cdot\frac{\detO}{\ONS{++}{0}-\ONS{+-}{0}}
  \label{eq:ONS02DMS3}
\end{align}

Combining Eq.~\ref{eq:ONS02DMS1}-\ref{eq:ONS02DMS3} and Eq.~\ref{eq:exp2ONS0},
we arrived at the same relation between $\mathfrak{D}_{ij}$ and experimental
measurables, as is studied in Ref.~\citenum{2015_BalsaraNewman} and\citenum{2018_VillaluengaPeskoEtAl}, which proves
the consistency between different descriptions of transport coefficients.
\begin{align}
  \mathfrak{D}_{+0} &= \frac{c_{0}D_{\mathrm{salt}}}{2c_{T}\tn{-}{0}\left(\tfac\right)}\\
  \mathfrak{D}_{+-} &= \left[\frac{c_{T}F^{2}}{\sigma RT}
                      -\frac{2\tn{+}{0}\tn{-}{0}\left(\tfac\right)}{cD}\right]^{-1}\\
  \mathfrak{D}_{-0} &= \frac{c_{0}D_{\mathrm{salt}}}{2c_{T}\tn{+}{0}\left(\tfac\right)}
\end{align}

\section{List of symbols}

\begin{tabular}{l l} 
$a_{i}^{\mathrm{S}}$ & weighing factor of species $i$ under reference frame $\mathrm{S}$ \\
$A_{ij}^{\mathrm{RS}}$ & matrix elements to convert the independent fluxes from reference frame $\mathrm{S}$ to $\mathrm{R}$\\
$c$ & salt concentration \\
$c_i$ & molar concentration of species $i$ \\
$c_T$ & total solution concentration  \\
$D_{\mathrm{salt}}$ & salt diffusion coefficient \\
$\mathfrak{D}_{ij}$ & Maxwell-Stefan diffusion coefficient for interaction of species $i$ and $j$ \\
$F$ & Faraday's constant \\
$\textbf{J}_{i}^{\mathrm{S}}$ & flux of species $i$ under reference frame $\mathrm{S}$ \\
$k_\mathrm{B}$ & Boltzmann constant \\
$K_{ij}$ & friction coefficient for interaction of species $i$ and $j$ \\
$m$ & molality \\
$M_{ij}$ & modified friction coefficient for interaction of species $i$ and $j$ \\
$M_{\mathrm{i}}$ & molecular weight of species $i$ \\
$n_{\mathrm{ion}}$ & number of ions (cation or anion)\\
$N_{\mathrm{A}}$ & Avogadro constant \\
$q_i$ & formal charge of species $i$ \\
$r$ & salt concentration ratio [Li/EO] \\
$\textbf{r}_i^\mathrm{S}$ & summed coordinate vector of species $i$ under reference frame $\mathrm{S}$ \\
$R$ & gas constant \\
$t$ & time interval \\
$t_{i}^{\mathrm{S}}$ & transference number of species $i$ under reference frame $\mathrm{S}$ \\
$T$ & temperature \\
$\textbf{v}_i$ & mean velocity of species $i$ \\
$\textbf{X}_{i}$ & external driving force on species $i$ \\
$z_i$ & charge number of species $i$ \\
\end{tabular}

\begin{tabular}{l l} 

$\gamma_{\pm}$ & mean molal activity coefficient of the salt \\
$\delta_{ij}$ & Kronecker delta function for interaction of species $i$ and $j$\\
$\mu_i$ & mobility of species $i$ \\
$\rho$ & density \\
$\rho_i$ & mass concentration of species $i$ \\
$\sigma$ & total ionic conductivity \\
$\beta$ & inverse temperature \\
$\omega_{i}$ & mass fraction of species $i$  \\
$\Omega_{ij}^{\mathrm{S}}$ & Onsager coefficient for interaction of species $i$ and $j$ under reference frame $\mathrm{S}$ \\
\end{tabular}

\newpage
\providecommand{\latin}[1]{#1}
\makeatletter
\providecommand{\doi}
  {\begingroup\let\do\@makeother\dospecials
  \catcode`\{=1 \catcode`\}=2 \doi@aux}
\providecommand{\doi@aux}[1]{\endgroup\texttt{#1}}
\makeatother
\providecommand*\mcitethebibliography{\thebibliography}
\csname @ifundefined\endcsname{endmcitethebibliography}
  {\let\endmcitethebibliography\endthebibliography}{}

%\bibliography{./references.bib}